\documentclass[%
reprint
superscriptaddress,
amsmath,amssymb,
aps,
prb,
twocolumn
]{revtex4-2}

\usepackage{graphicx}% Include figure files
\usepackage{dcolumn}% Align table columns on decimal point
\usepackage{bm}% bold math
\usepackage{multirow}
\usepackage{float}
\usepackage{overpic}
\usepackage{subcaption}
\usepackage{caption}
\usepackage{makecell}
\usepackage{oplotsymbl}
\usepackage{todonotes}
\usepackage{hyperref}% add hypertext capabilities
\usepackage{color}
\usepackage{graphpap}
\usepackage{MnSymbol}

\begin{document}

\preprint{APS/123-QED}

\title{First-principles analysis of the effect of magnetic states on the oxygen \\ vacancy formation energy in doped La$_{0.5}$Sr$_{0.5}$CoO$_3$ perovskite}

\author{Wei Wei}
\email{wei.wei@enas.fraunhofer.de} %\email{wei.wei@zfm.tu-chemnitz.de}
\affiliation{Center for Microtechnologies, Chemnitz University of Technology, Chemnitz 09107 Saxony, Germany}
\affiliation{Fraunhofer Institute for Electronic Nano Systems, Chemnitz 09126 Saxony, Germany}

\author{Florian Fuchs}
\affiliation{Center for Microtechnologies, Chemnitz University of Technology, Chemnitz 09107 Saxony, Germany}
\affiliation{Fraunhofer Institute for Electronic Nano Systems, Chemnitz 09126 Saxony, Germany}

\author{Andreas Zienert}
\affiliation{Center for Microtechnologies, Chemnitz University of Technology, Chemnitz 09107 Saxony, Germany}
\affiliation{Fraunhofer Institute for Electronic Nano Systems, Chemnitz 09126 Saxony, Germany}

\author{Xiao Hu}
\affiliation{Center for Microtechnologies, Chemnitz University of Technology, Chemnitz 09107 Saxony, Germany}
\affiliation{Fraunhofer Institute for Electronic Nano Systems, Chemnitz 09126 Saxony, Germany}

\author{Jörg Schuster}
\affiliation{Center for Microtechnologies, Chemnitz University of Technology, Chemnitz 09107 Saxony, Germany}
\affiliation{Fraunhofer Institute for Electronic Nano Systems, Chemnitz 09126 Saxony, Germany}

\date{\today}

\begin{abstract}

Oxygen vacancies are critical for determining the electrochemical performance of fast oxygen ion conductors. The perovskite La$_{0.5}$Sr$_{0.5}$CoO$_3$, known for its excellent mixed ionic-electronic conduction, has attracted significant attention due to its favorable vacancy characteristics. In this study, we employ first-principles calculations to systematically investigate the impact of 3$d$ transition-metal doping on the oxygen vacancy formation energies in the perovskite. Two magnetic states, namely the ferromagnetic and paramagnetic states, are considered in our models to capture the influence of magnetic effects on oxygen vacancy energetics. Our results reveal that the oxygen vacancy formation energies are strongly dependent on both the dopant species and the magnetic state. Notably, the magnetic states alter the vacancy formation energy in a dopant-specific manner due to double exchange interactions, indicating that relying solely on the ferromagnetic ground state may result in misleading trends in doping behavior. These findings emphasise the importance of accounting for magnetic effects when investigating oxygen vacancy properties in perovskite oxides. 

\end{abstract}

\maketitle

\section{Introduction\protect} \label{Introduction}

Defect-driven transport phenomena are fundamental to the operation of fast ion-conducting solid electrodes and electrolytes \cite{poletayev2022defect}. During operation, intrinsic or extrinsic defects serve as the primary pathways for ionic migration. Among these defects, the critical role of oxygen vacancies extends to a broad class of fast-ion conductors, where their presence and mobility directly impact the efficiency of oxygen-ion transport \cite{shao2004high, steele2001materials}. Materials that accommodate high concentrations of oxygen vacancies, such as fluorite and perovskite crystals, often exhibit enhanced ionic conductivity and redox behavior, which are vital for applications including solid oxide fuel cells (SOFCs) and oxygen separation membranes \cite{mogensen2004factors,orera2010new}. Within this context, perovskite crystals have garnered significant attention due to their tunable defect chemistry and fast oxygen-ion conduction. 

The perovskite framework, namely the compound of ABO$_3$, exhibits enhanced flexibility among the family of metal oxides \cite{sun2021recent}. This superior adaptability facilitates a broader range of ionic substitutions and the accommodation of various structural defects. Among them, the cobalt-based perovskites are a prominent category due to their exceptional oxygen diffusivity and oxygen surface exchange coefficient \cite{kilner2000fast}. Various studies of A-position and B-position occupants have been added to such materials to accommodate a wide range of operating conditions. The inherent tunability of their B-site cation allows for precise engineering of oxygen vacancy formation and ionic transport properties \cite{skinner2003oxygen}. For instance, partial substitution of Co with other transition metal (TM) elements  has been extensively explored to enhance stability under operational conditions or optimize ionic conductivity \cite{lee2015oxygen}. 

Despite these advances, open questions remain concerning the interplay between dopant chemistry and oxygen vacancy behavior, particularly under different manufacturing methods. Previous experimental studies have developed strategies such as nanostructuring, surface modification, and composite designs to improve the oxygen reduction reaction activity and durability of perovskite cathodes \cite{wachsman2011lowering,adler2004factors}. However, these methods frequently depend on the synergistic effect of multiple factors, resulting in the potential for contradictory outcomes for the same dopant elements. For example, Cheng $et$~$al.$ reported that Fe doping increases the concentration of lattice oxygen vacancies, thereby enhancing the performance of La$_{1-x}$Sr$_x$CoO$_{3-\delta}$ \cite{cheng2022bifunctional}. In contrast, Mantzavinos $et$~$al.$~observed that increasing Fe content leads to a decrease in oxygen vacancy concentration \cite{mantzavinos2000oxygen}. Consequently, theoretical calculations have been sought to elucidate the intrinsic effect of doping elements on the nature of oxygen vacancies in the perovskite \cite{ingavale2024strategic}.

Extensive theoretical studies have been conducted to investigate oxygen vacancy formation in cobalt-based perovskites \cite{ingavale2024strategic,jia2024alternative,li2023first}. However, these studies often focus on single-element doping effects and assume an idealized ferromagnetic (FM) ground state, neglecting the paramagnetic (PM) states that prevail at elevated temperatures during device operation \cite{adler2004factors}. This oversight limits the predictive accuracy and rationality of computational models. For instance, Jia $et$~$al.$~demonstrated that Fe doping of cobalt-based perovskites decreases oxygen vacancy formation energy in the PM state \cite{jia2021effects}. This variation underscores the necessity of incorporating both magnetic states into theoretical investigations to accurately predict dopants behavior in operational environments.

In this work, we systematically investigate the impact of substituting Co with Mn, Fe, Ni, and Cu on the energetics of oxygen vacancy formation in La$_{0.5}$Sr$_{0.5}$CoO$_3$ (LSC), considering both FM and PM states. Using density functional theory (DFT), we evaluate the interplay among dopant electronic configurations, magnetic order/disorder, and oxygen vacancy stability. These findings provide key insights for designing next-generation perovskite SOFC cathodes with tailored oxygen vacancy concentrations and enhanced electrochemical performance.

The paper is organized as follows: In Sec. \ref{Theoretical_approach} we introduce the computational setup and the definition of the formation energy of oxygen vacancies in LSC. In Sec. \ref{results}, we present the results on stability of oxygen vacancies by different magnetic states, and its influence on the electronic structure. In Sec. \ref{discussion}, we discuss our results in view of other findings in the literature, and we summarize and conclude in Sec. \ref{summary}.

\section{Theoretical approach} \label{Theoretical_approach}

\subsection{Crystal structure of LSC} \label{crystal}

\begin{figure}
	\centering
	\begin{overpic}[width=0.48\textwidth]{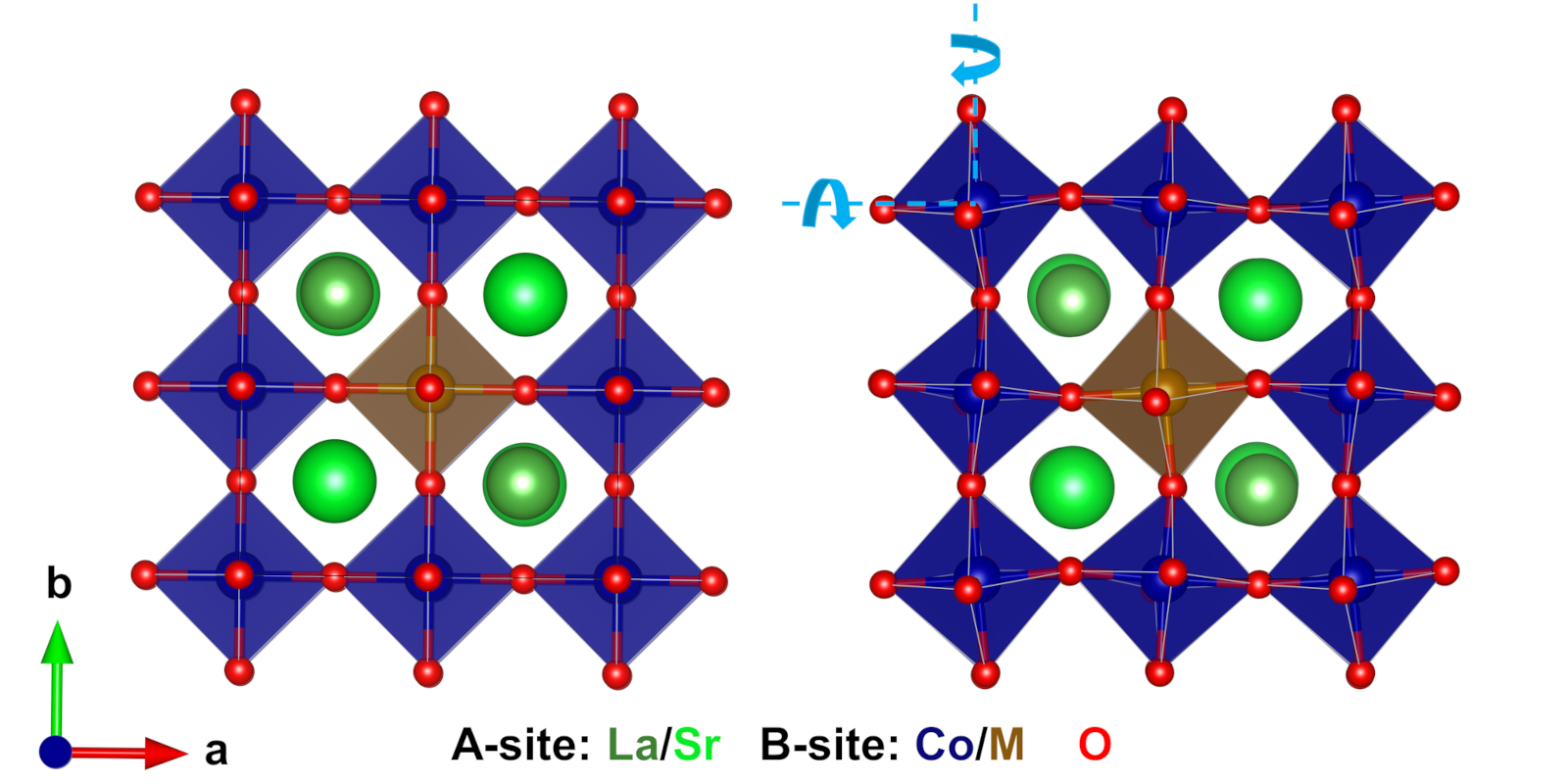}
		\put(3,45){(a)}
		\put(50,45){(b)}
	\end{overpic}
	\caption{Atomic structures of doped LSC. (a) The ordered perovskite structure with the dopant M substituting for Co at a B-site. (b) The distorted structure induced by the J-T effect. The light blue arrows indicate the rotation of the B-O octahera. La/Sr atoms are shown in dark/light green, Co atoms in blue, O atoms in red, and the dopant site by M is marked in dark gold. }
	\label{fig1}
\end{figure}

In earlier studies, the parent LSC phase was considered to adopt a cubic perovskite structure (space group $Pm\bar{3}m$) under operating conditions and room temperature \cite{senaris1995magnetic,bhide1975itinerant}. The structure consists of corner-sharing CoO$_6^{2/3-}$ octahedra, with La$^{3+}$ and Sr$^{2+}$ ions randomly occupying the A-site, creating a disordered cationic sublattice. Investigations on similar perovskites have further revealed that the energies associated with different La/Sr configurations are very close, allowing the use of isotropic supercells to reliably capture the average structural and electronic properties of these materials \cite{meng2015synergistic,pavone2014first}. A perfect cubic model with an ordered La/Sr configuration is depicted in Figure~\ref{fig1}a. The lattice constant of this cubic model is 3.85~\AA~by DFT, which is in good agreement with the experimental values of 3.83~\AA~\cite{walter2017ion,woicik2011effect}. However, more recent studies indicate that under operating conditions, particularly in thin films, LSC does not maintain a perfect cubic symmetry but rather exhibits a pseudo-cubic structure characterized by distortions in the Co-O octahedra \cite{kubicek2013tensile,jia2024alternative,kamecki2021improvement}. 

The pseudo-cubic structure of LSC with distorted Co-O octahedra can be mainly attributed to the Jahn-Teller (J-T) effect \cite{mizokawa1995unrestricted}. The J-T effect arises from electronic degeneracies in the Co $d$ orbitals, which drive a spontaneous symmetry breaking in the CoO$_6$ octahedra to lower the energy of the system \cite{rata2008strain,cai2011surface}. Such structural modifications not only alter the local lattice geometry but also have profound implications on the electronic structure and oxygen vacancy formation energetics, thereby influencing the electrochemical performance of the material \cite{louca1999correlation}. Therefore, we have incorporated the structural distortion into our model, as illustrated in Figure~\ref{fig1}b. 

In our models, we replace one of the Co atoms with the aforementioned TM elements. At such low doping concentrations, the crystal retains its pseudo-cubic structure, and the lattice constant does not change significantly \cite{mandal2023electrocatalytic}. Therefore, we use the same size of the simulation box for all the models.

\subsection{Magnetic order calculations} \label{magnetic}

The magnetic properties of LSC are governed by the complex interplay between Co$^{3+}$/Co$^{4+}$ spin states and oxygen vacancy distributions \cite{baskar2008high}. At temperatures lower than 250 K, LSC exhibits FM ordering due to double exchange between two Co ions through the intermediate shared O atom \cite{senaris1995magnetic,bhide1975itinerant}. However, at most operating temperatures, thermal fluctuations disrupt the long-range FM order, leading to a PM state with short-range spin correlations \cite{baskar2008high}.

\begin{figure}
	\centering
	\begin{overpic}[width=0.48\textwidth]{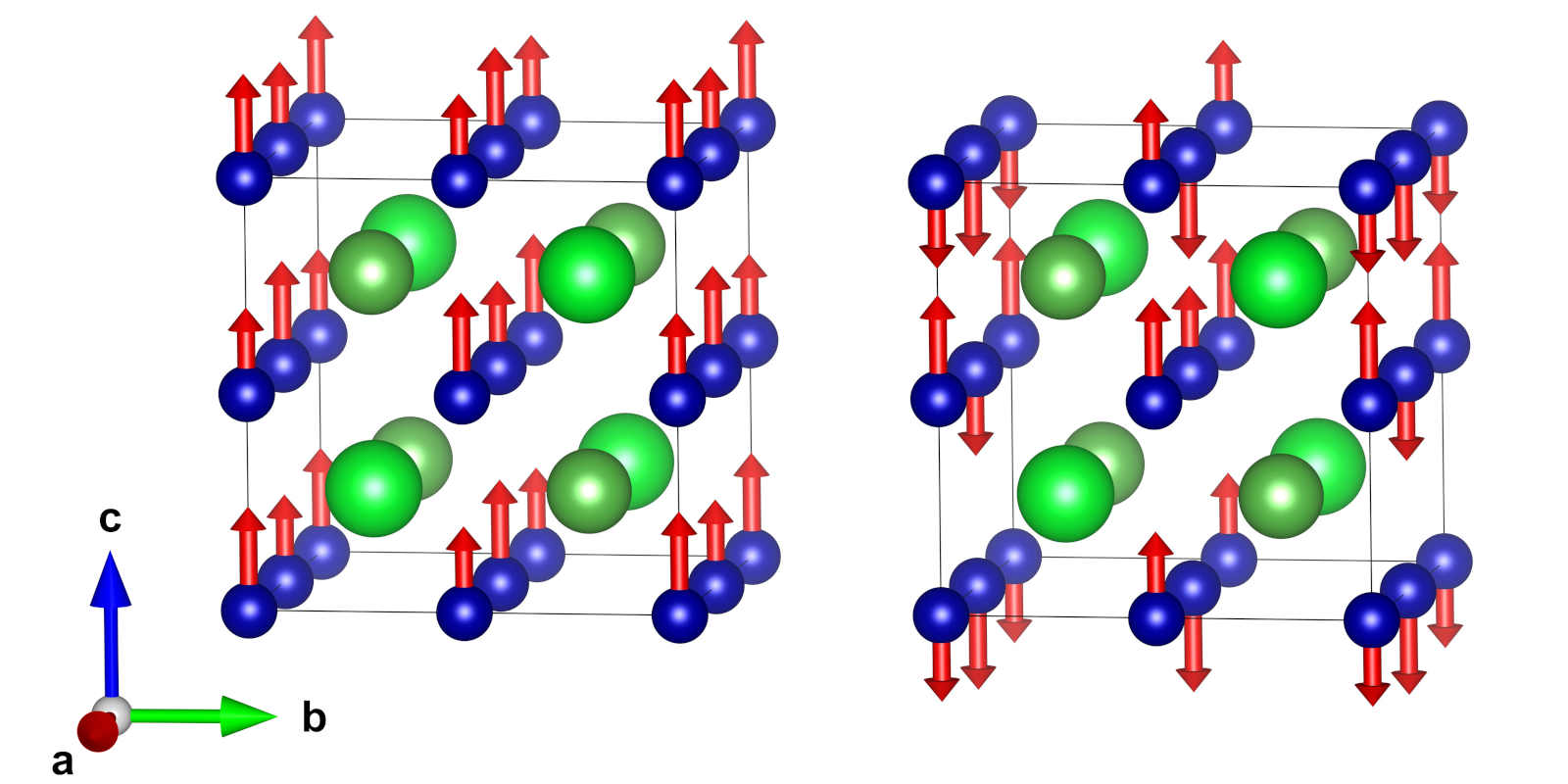}
		\put(3,45){(a)}
		\put(52,45){(b)}
	\end{overpic}
%	\begin{overpic}[width=0.22\textwidth]{./figures/fig2b.png}
%		\put(-5,95){(b)}
%	\end{overpic}
	\caption{Magnetic configurations of LSC: (a) the FM state, and (b) one of the PM state configurations. The magnetic moments of TM elements are represented by red arrows. To improve clarity, oxygen atoms are not shown. }
	\label{fig2}
\end{figure}

Simulating the exact noncollinear magnetic configurations in the PM state is computationally expensive due to the size of the supercell and the need for extensive statistical sampling. Instead, we employ the Magnetic Sampling Method (MSM) \cite{alling2010effect}, which simulates PM states using the average of a certain mount of collinear magnetic configurations, as shown in Figure \ref{fig2}. In each of the samples, the total magnetic moment is set to 0~$\mu_B$. The MSM approach has been validated against more rigorous methods (e.g., DLM-CPA) for transition metal oxides, yielding comparable accuracy in predicting magnetic and electronic properties \cite{alling2010effect}.

In this work, the MSM implementation involves generating 20 distinct collinear spin configurations for our models. This approach captures the essential physics of the PM state while maintaining computational feasibility, as demonstrated by its successful application to similar transition metal oxides systems \cite{merkel2024probing,yoon2021effects,golosova2017magnetic}.

\subsection{Computational details} \label{computational}
 
The calculations are carried out with the Vienna ab initio simulation package \cite{kresse1996efficiency} employing projector-augmented waves  \cite{blochl1994projector} and the spin-polarized Perdew-Burke-Ernzerhof \cite{perdew1996generalized} generalized gradient approximation. The interaction between valence electrons and ionic cores is modeled by treating the 5$s$, 5$p$, and 6$s$ orbitals of La, the 4$s$, 4$p$, and 5$s$ orbitals of Sr, the 3$s$, 3$p$, 3$d$, and 4$s$ orbitals of TM elements, and the 2$s$ and 2$p$ orbitals of O as valence electrons. To avoid the self-interaction errors that occur in the standard DFT for strongly correlated electronic systems, we employ the DFT+U method \cite{dudarev1998electron} accounting for the on-site Coulomb interaction in the localized $d$ orbital. 

In our calculations, an energy cutoff of 600~eV is used for the plane-wave basis. Total energy differences and forces on atoms for all structural degrees of freedom are converged within $1 \times10^{-5}$~eV and $5 \times10^{-2}$~eV/\AA, respectively. The Brillouin-zone integrals are sampled by 4$\times$4$\times$4 Monkhorst-Pack $k$-point grids \cite{monkhorst1976special} with a Gaussian smearing of $1\times10^{-3}$~eV for
the 2$\times$2$\times$2 supercell models containing 40 atoms with composition La$_{0.5}$Sr$_{0.5}$Co$_{0.875}$M$_{0.125}$O$_{3}$ (M = Mn, Fe, Co, Ni, Cu). Structural relaxations for these systems are carried out using the Fast Inertial Relaxation
Engine (FIRE) algorithm \cite{bitzek2006structural}. The specific U values employed are 4.0~eV for Mn, Cu and Fe, 3.3~eV for Co, and 6.4~eV for Ni \cite{wang2006oxidation}. Defect formation energies of TM dopants located at B sites in the perovskite crystal are calculated as 
\begin{equation}
	E_{\rm TM}^{\rm doping}=\frac{E_{\rm (La/Sr)Co_{1-y}M_{y}O_{3}} - E_{\rm LSC}^{\rm bulk} - y (E_{\rm M}^{\rm metal}+E_{\rm Co}^{\rm metal})}{y}
\end{equation}
where $E_{\rm (La/Sr)Co_{1-y}M_{y}O_{3}}$ is the total energy per formula unit of the doped perovskite crystal, $E_{\rm LSC}^{\rm bulk}$ is the energy per formula unit of the pure LSC reference crystal, $E_{\rm M}^{\rm metal}$ and $E_{\rm Co}^{\rm metal}$ are the energy per atom in the most stable phase of the elemental metals, and $y$ is the proportion of dopant per formula unit of the perovskite crystal. 

Similarly, the formation energy of an oxygen vacancy in the perovskite crystal is calculated as
\begin{equation}
	E_{\rm V}^{\rm O}=E_{\rm sys}^{\rm vac} - E_{\rm sys}^{\rm bulk} + \frac{1}{2} E_{\rm O_2}
\end{equation}
where $E_{\rm sys}^{\rm vac}$ is the total energy of the (doped or non-doped) perovskite crystal with an oxygen vacancy, $E_{\rm sys}^{\rm bulk}$ is the total energy of the same crystal without oxygen vacancies, and $E_{\rm O_2}$ is the energy of an oxygen molecule. In our calculations, we consider a single oxygen vacancy by removing one O atom from a total of 24 O atoms in our model.

\section{Results} \label{results}

\subsection{Structural distortion of LSC}

\begin{figure}
	\centering
	\begin{overpic}[width=0.45\textwidth]{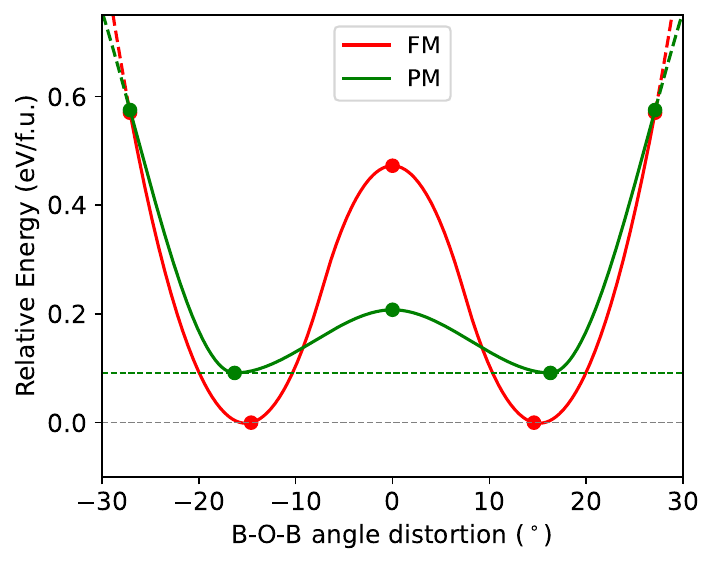}
	\end{overpic}
	\caption{Relative energy per formula unit as a function of B-O-B angle distortion in LSC for FM (red) and PM (green) configurations. The cubic structure corresponds to the local energy maximum at 0° distortion, indicating its instability due to the Jahn-Teller effect. All energies are referenced to the lowest energy point in the FM configuration, marked as 0~eV on the vertical axis.}
	\label{fig3}
\end{figure}

To elucidate the energy landscape associated with distortions of CoO$_6$ octahedra in LSC, we investigated the relative energy as a function of the B-O-B bond angle distortion. The J-T effect plays a crucial role in the distortions, leading to the instability of the ordered cubic structure, which manifests as a saddle point in the potential energy surface. As demonstrated in Figure~\ref{fig3}, we compared the potential energy surface of two distinct magnetic configurations: FM and PM states. In our results, all energies are referenced to the lowest energy point in the FM configuration, which is set to 0~eV. The energy profiles exhibit a characteristic double-well potential, with minima corresponding to distorted structures and a local energy maximum at the undistorted cubic phase (0° distortion). Deviations from 0° represent increasing distortion, with positive and negative angles indicating different distortion directions.

The stability of different magnetic configurations varies with structural distortions. The FM configuration exhibits its minimum energy at a distorted angle of ±15°. While, the PM configuration shows a higher energy minimum at a similar distortion angle, which is ±16°, with an energy difference of 0.09~eV per formula unit relative to the FM state. At the 0° distortion, the PM state is more stable than the FM state, with the PM energy being 0.3~eV lower than that of the FM configuration. These observations indicate that while both FM and PM states favor distorted structures over the cubic phase, the FM state stabilizes the distorted phases more effectively, whereas the PM state provides greater stability to the ordered cubic structure.

\begin{figure}
	\centering
	\begin{overpic}[width=0.48\textwidth]{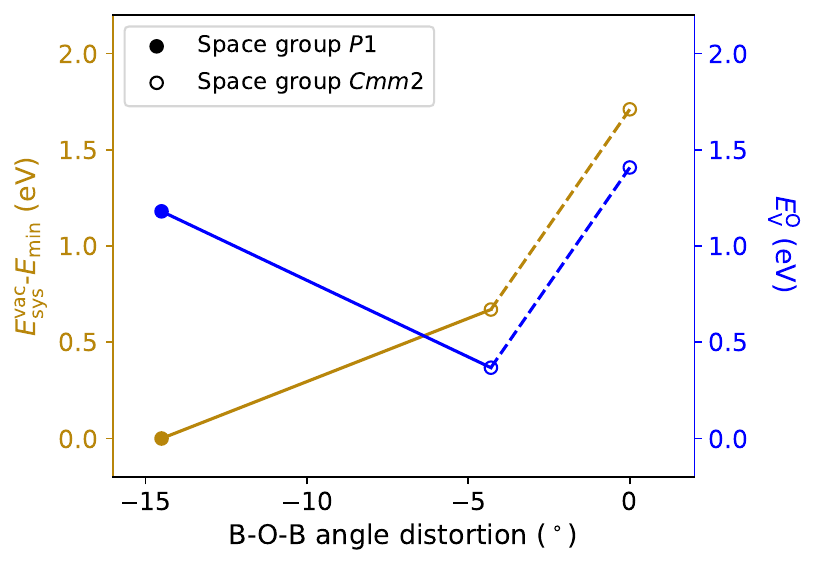}
	\end{overpic}
	\caption{Energy analysis of vacancy-containing LSC structures as a function of B-O-B angle distortion. The dark yellow curve represents the total system energy $E \rm _{sys}^{vac}$, referenced to the lowest energy configuration $E \rm _{min}$. The blue curve shows the oxygen vacancy formation energy $E \rm _V^O$. The solid and open markers indicate different symmetry of the crystal.}
	\label{fig4}
\end{figure}

To further investigate the role of structural distortion in the presence of oxygen vacancies, we analyze the total energy of vacancy-containing systems as a function of B-O-B bond angle distortion, as shown in Figure~\ref{fig4}. We compute the total energy in three distinct ways: first, the ordered structure without relaxation (space group $Cmm2$), second, the ordered structure with relaxation (space group $Cmm2$), and third, the distorted structure with relaxation in no symmetry (space group $P1$). In the three structures, the distortion increases from the first to the last. From the data points, we observe that the total energy decreases as the distortion increases, indicating that structural distortions stabilize the system in the presence of oxygen vacancies.

In addition to the system energy, we illustrate the oxygen vacancy formation energies in Figure~\ref{fig4}. For the distorted structure, the vacancy formation energy is approximately 1.2~eV. On the other hand, when using the pristine ordered structure (space group $Fm\bar{3}m$) as a reference, the vacancy formation energy in a symmetry-preserved system (space group $Cmm2$) is notably reduced to around 0.4~eV, demonstrating the strong influence of structural distortions on defect energetics. Furthermore, a non-relaxed ordered structure yields a vacancy formation energy of approximately 1.4~eV, similar to the value obtained for the distorted system. However, this result is unrealistic since oxygen vacancy defects inevitably induce local structural relaxations. Therefore, in subsequent calculations, we consider only the vacancy formation energy obtained from the structurally distorted configurations.

\subsection{Oxygen vacancies in doped LSC} \label{results_b}

In order to study the effects of dopant elements on the formation energy of oxygen vacancies, an initial investigation of the formation energy of substituting a single Co atom with TM dopants (Mn, Fe, Ni, Cu) in the pristine LSC crystal is conducted. We then calculate the formation energy of oxygen vacancies in the investigated material systems.

\begin{figure}
	\centering
	\begin{overpic}[width=0.45\textwidth]{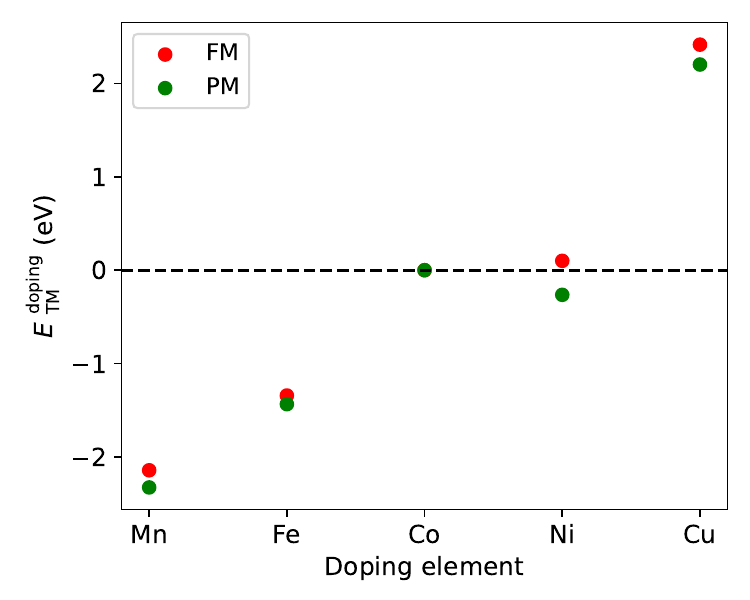}
	\end{overpic}
	\caption{Doping energy of pristine LSC with transition metal dopants (Mn, Fe, Ni, Cu) substituting a single Co atom. Red and green circles denote ferromagnetic (FM) and paramagnetic (PM) configurations, respectively. In both magnetic states, the formation energy of pure LSC, indicated by Co, was set to be zero as the reference.}
	\label{fig5}
\end{figure}

Figure~\ref{fig5} illustrates the TM-doping energies under FM and PM states. A clear correlation is observed between the TM doping energy trends in FM state, suggesting that the stability of doped LSC is governed by the intrinsic oxidation energies of the dopants. This trend is similar to the reported oxide formation energies of the dopant elements \cite{wang2006oxidation}. The doping energies in the PM state exhibit a slight reduction ($0.1$--$0.3$~eV) compared to those in the FM state across all dopants. However, this difference does not alter the overall energetic trend, which follows the sequence Mn $\less$ Fe $\less$ Co $\less$ Cu. The consistent trend implies that magnetic ordering barely affects the relative stability of dopants.   

The only outlier is Ni dopant. The Ni dopant maintains the doping energy trend in the FM configuration, consistent with its strong tendency to form stable oxides. However, its energy reduction in the PM state ($\sim 0.3$~eV) is slightly larger than that of other dopants, leading to enhanced stability than pure LSC. This phenomenon is analogous to observations in other Ni-based perovskite systems, where the material has a tendency to stabilize PM or anti-FM states at low temperature \cite{zunger2021understanding}.   

\begin{figure}
	\centering
	\begin{overpic}[width=0.45\textwidth]{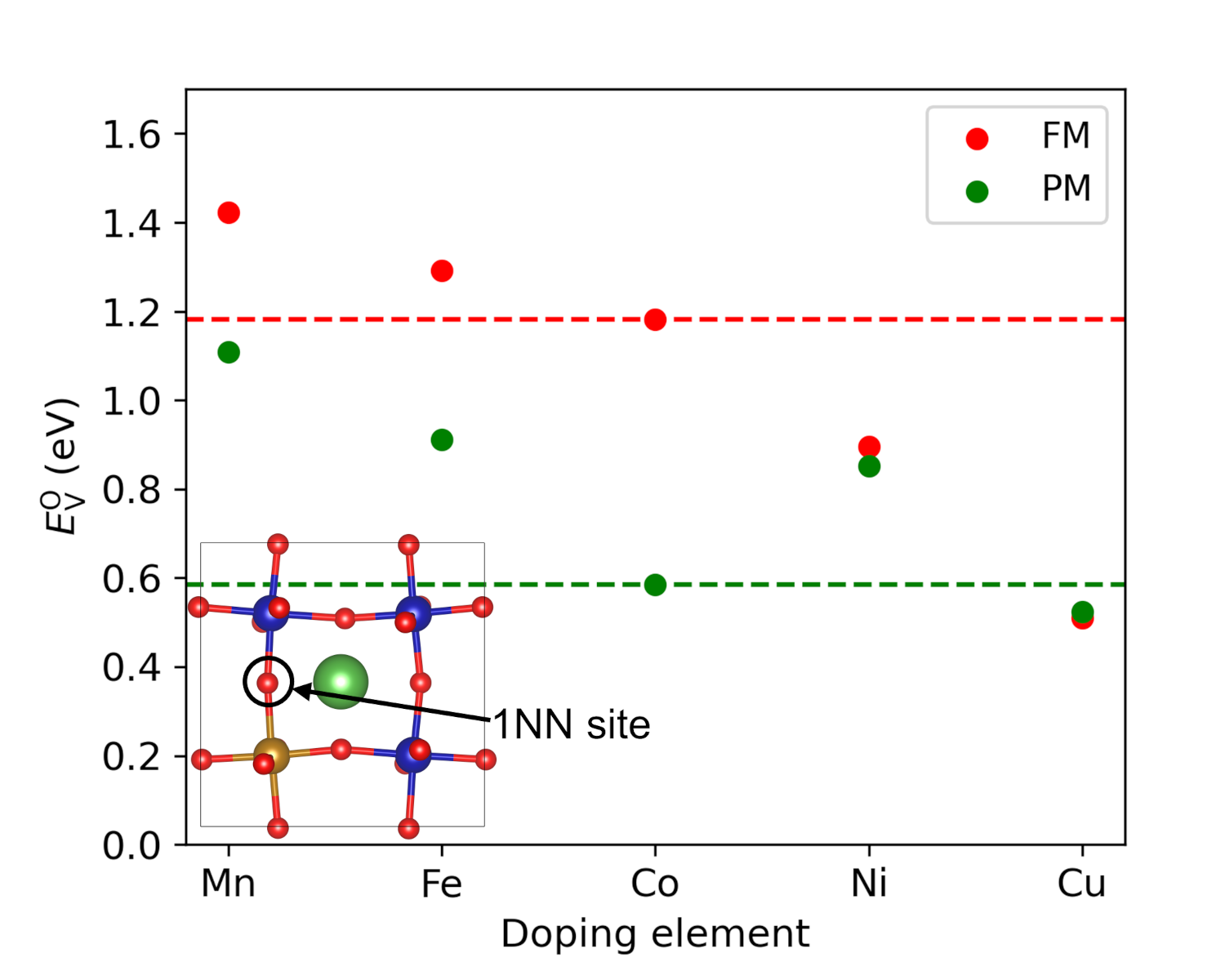}
		\put(0,75){(a)}
	\end{overpic}
	\begin{overpic}[width=0.45\textwidth]{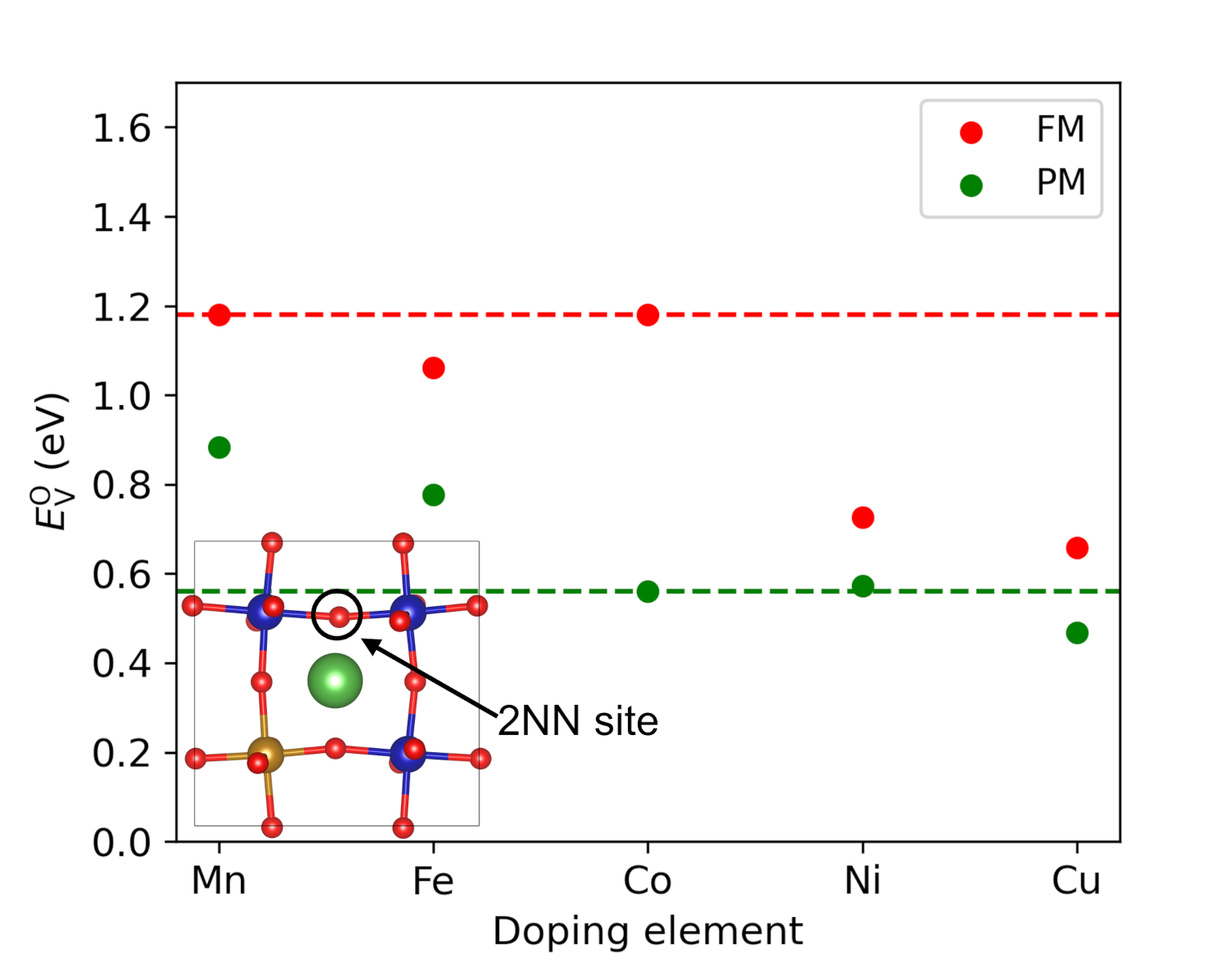}
		\put(0,75){(b)}
	\end{overpic}
	\caption{Oxygen vacancy formation energy (a) at the 1NN and (b) at the 2NN sites of the doped LSC. FM and PM configurations are shown in red and green, respectively. For better comparison, the dashed lines mark $E \rm _V^O$ for the Co.}
	\label{fig6}
\end{figure}

We then investigate $E \rm _V^O$ in doped LSC depending on the magnetic states. Figure~\ref{fig6} compares $E \rm _V^O$ for first-nearest-neighbor (1NN) and second-nearest-neighbor (2NN) vacancy sites relative to the dopant site, respectively.

%We first discuss the vacancy formation energies relative to the undoped LSC. The two oxygen vacancies demonstrate slightly different trends in vacancy formation energies under the TM doping. The doping elements have been observed to affect not only the O atoms with which they form chemical bonds, but also the other oxygen atoms by affecting the neighboring Co atoms. In the FM state, the $E \rm _V^O$ of 1NN vacancy are associated with the oxidizability of the TM, with energies ranked from Mn to Cu dopants from high to low. However, for 2NN vacancy, their formation energies exhibit a decrease at Mn, Fe, and Ni dopants and an increase at Cu dopant. For the PM states, the different positions also had some effects on the vacancy formation energy, while these effects did not shift the energetic order of the vacancy formation energy at different dopants.

First, we investigate the effect of vacancy position on their formation energies. The doping elements affect not only the O atoms with which they form chemical bonds, but also the other O atoms by affecting the neighboring Co atoms. In the presence of TM dopants, differences in the formation energies of two different oxygen vacancy sites have been observed. These variations can even alter the relative stability (compare the $E \rm _V^O$ in pure LSC) of the vacancies. Therefore, it is essential to consider both vacancy sites to understand the dopant-induced effects on the stability of the vacancies.

%Except the position of oxygen vacancies, it has also been demonstrated that alterations in the magnetic ordering exert a substantial influence on the oxygen vacancy formation energy. 
Next, we compare the vacancy formation energy between the FM and PM state. The PM state lowers $E \rm _V^O$ (compared to FM) for both 1NN and 2NN sites in most of the situations, which suggests that the PM state weakens the strength of TM-O bonds. This weakening is most pronounced in the pure LSC, resulting in an energy difference of 0.6~eV (difference between the dashed lines in Figure~\ref{fig6}). For Fe and Mn dopants, this energy difference ranges from 0.3~eV to 0.4~eV. However, the situation becomes a bit more complicated for Ni and Cu doping. The vacancy formation energy remains largely unaltered at the 1NN vacancy nearest the doping site. In contrast, the formation energy of the 2NN vacancy is reduced by around 0.2~eV.

Vacancy positions and magnetic states together create even more complex variations in oxygen vacancy stability, with the effects of Fe and Ni dopants being particularly pronounced. For the 1NN vacancy, the Fe-doped FM state exhibits a higher vacancy formation energy compared to the LSC. However, this tendency is reversed at the 2NN site. In contrast, for the PM state, the Fe-doped system exhibits higher vacancy formation energies than the LSC for both vacancy positions. For Ni doping, the vacancy formation energy is consistently lower in the FM state than in the LSC, while in the PM state, the vacancy formation energy is always higher than its formation energy in the LSC. Consequently, the energy change induced by the PM magnetic state qualitatively affects stability of the oxygen vacancy in the doped system.

\begin{figure}
	\centering
	\begin{overpic}[width=0.45\textwidth]{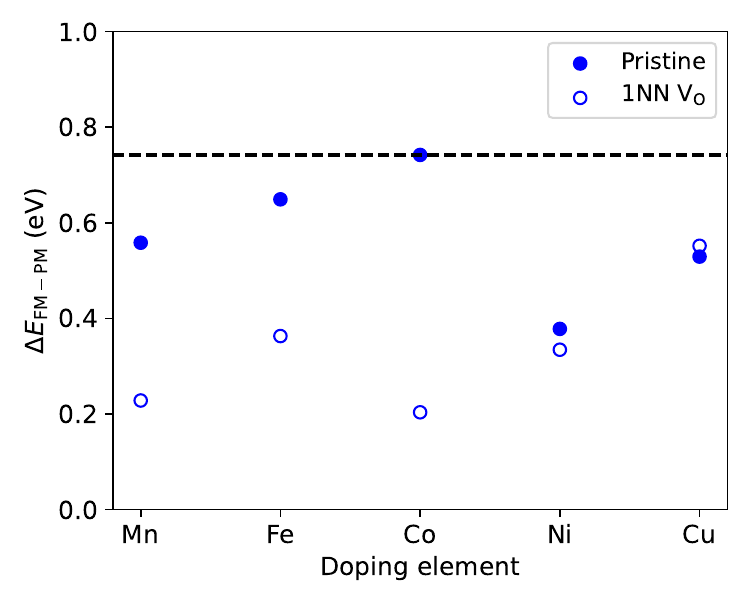}
	\end{overpic}
	\caption{Energy difference between PM and FM states ($\Delta E{\rm _{FM-PM}}$) for pristine crystals and crystals with 1NN V$_{\text{O}}$, denoted as solid and open markers, respectively.}
	\label{fig7}
\end{figure}

To further elucidate the interplay between defects and magnetic states, Figure~\ref{fig7} quantifies the energy difference between the two magnetic states ($\Delta E{\rm _{FM-PM}}$) for the crystals with and without oxygen vacancy. The results demonstrate an absence of regularity in the apparent changes in $\Delta E{\rm _{FM-PM}}$.

Specifically, the introduction of an oxygen vacancy alters $\Delta E{\rm _{FM-PM}}$ in a dopant-dependent manner, suggesting that local structural and electronic rearrangements around the vacancy site significantly influence spin alignment preferences. The most pronounced change in $\Delta E_{\text{FM-PM}}$ due to oxygen vacancy formation is observed in pristine LSC, where the energy difference decreases by approximately 0.5~eV. In contrast, Mn- and Fe-doped systems exhibit a more moderate reduction in $\Delta E_{\text{FM-PM}}$, with values decreasing by around 0.35~eV. On the other hand, for Ni and Cu doping, the energy difference remains largely unchanged regardless of the presence or absence of an oxygen vacancy, indicating a weaker dependency between defect formation and magnetic stability in these systems.  

The results of this section illustrate the interplay among doping chemistry, oxygen vacancy defects, and magnetic order. However, these relationships do not exhibit a uniform trend. In particular, while our aim is to qualitatively elucidate the relationship between oxygen vacancy formation energy and magnetic order, significant variations are observed across different compositional systems. Attempts to correlate these variations with crystallographic parameters—such as TM-O bond lengths, bond angles, and ionic radii—were unsuccessful. Consequently, our calculations indicate that there is a complex interplay of different factors, and it is necessary to investigate the electronic structure of the systems.

\subsection{Electronic structure analysis}

In order to assess the doping-induced modifications in the electronic structure of LSC, we conduct a comparative analysis of the projected density of states (pDOS) for the doped crystals. The pDOS of crystals without $\text{V}_\text{O}$ are compared between the FM and PM states. For the sake of clarity, the contributions from La and Sr elements had been omitted from the figures.

\begin{figure}
	\centering
	\begin{overpic}[width=0.46\textwidth]{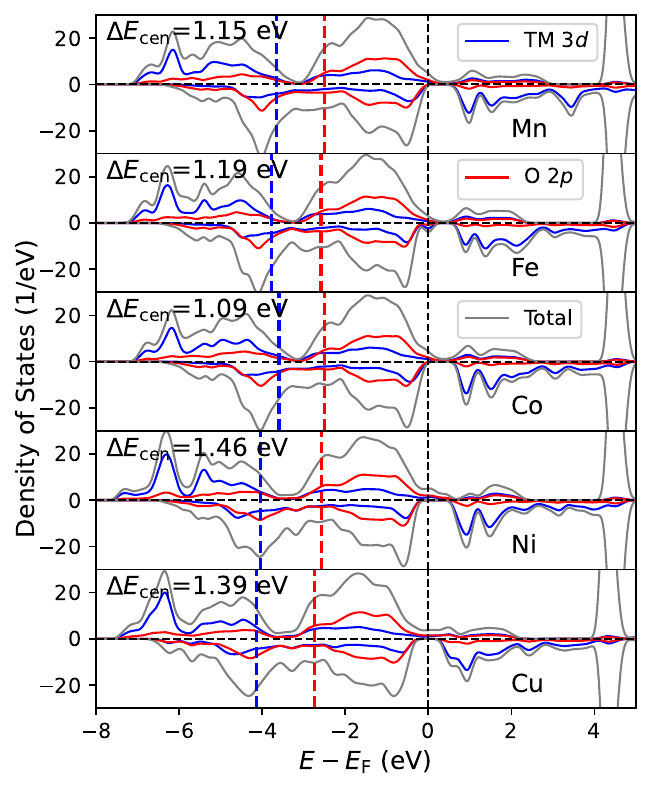}
		\put(0,95){(a)}
	\end{overpic}
	\begin{overpic}[width=0.46\textwidth]{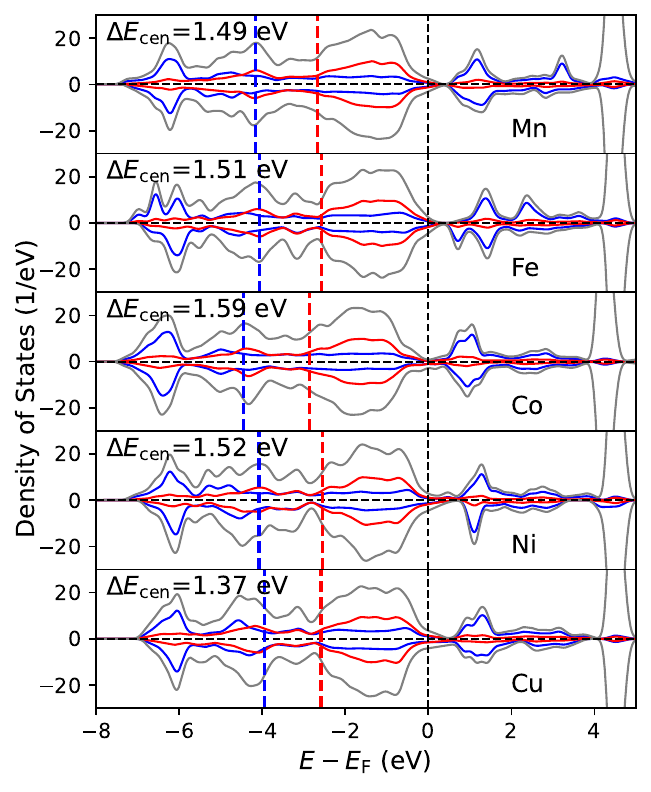}
		\put(0,95){(b)}
	\end{overpic}
	\caption{Projected density of states for doped LSC of (a) the FM state and (b) the PM state. Each panel displays the TM 3$d$ orbital (blue), the O 2$p$ orbital (red), and the total density of states (gray), with energies referenced to the Fermi level ($E \rm _F$). Spin-up and spin-down data are shown in each upper and lower panel, respectively. Vertical dashed lines denote the band centers of the occupied O 2$p$ (red) and TM 3$d$ (blue) states, and $\rm \Delta$$E$$\rm _{cen}$ is the energy difference between the two centers.}
	\label{fig8}
\end{figure}

Figure~\ref{fig8} illustrates the detailed information of the pDOS.  The band center, as the famous effective electronic descriptor \cite{hammer1995gold}, is also labeled in the figure. The band center is defined as the energy-weighted average of the orbital,~i.e., $\varepsilon_{\text{center}} = \frac{\int E \cdot D(E) \, dE}{\int D(E) \, dE}$ \cite{ando2021enhancement}. Across all dopants, the overall energy landscape remains qualitatively similar, indicating that the fundamental B-O bonding framework is preserved. This suggests that substituting different TM elements at the B-site does not introduce drastic modifications to the hybridization between O 2$p$ and TM 3$d$ orbitals. However, the energy difference between the center of the O 2$p$ band and TM 3$d$ band ($\rm \Delta$$E$$\rm _{cen}$) is affected by the dopants and the magnetic states. For the FM state, Mn and Fe dopants maintain a similar $\rm \Delta$$E$$\rm _{cen}$ to LSC, while the Ni and Cu dopants expand it by 0.37 and 0.30~eV, respectively. In contrast, Mn, Fe, and Ni dopants in the PM state show a similar $\rm \Delta$$E$$\rm _{cen}$, while the Cu dopant decreases the $\rm \Delta$$E$$\rm _{cen}$ around 0.2~eV.

Furthermore, the change in $\rm \Delta$$E$$\rm _{cen}$ between the two magnetic states is also related to the type of dopants. Specifically, as the magnetic state transitions from FM to PM state, the $\rm \Delta$$E$$\rm _{cen}$ increases by 0.5~eV within a pure LSC crystal. The increase for Mn and Fe dopants is slightly less than that for LSC, 0.34 and 0.32~eV, respectively. The Ni and Cu dopants exhibit a comparatively negligible change below 0.1~eV.

In summary, our pDOS analysis reveals that, while the overall energy landscapes of the TM 3$d$ and O 2$p$ states remain qualitatively similar across all dopants, indicating that the fundamental B–O bonding framework is preserved, the band centers exhibit modifications that are dopant- and magnetic state-dependent. These results underscore the complex interplay between dopant chemistry and magnetic states in modulating the electronic structure of LSC.

\section{Discussion} \label{discussion}

In this study, we presented the critical role of structural distortions in stabilizing the LSC system and modulating the formation energy of oxygen vacancies. Our findings are consistent with previous studies that have highlighted the significance of J-T distortions and octahedral tilting in transition metal oxides \cite{mizokawa1995unrestricted,balachandran2013interplay}. Specifically, we observe that the cubic phase of LSC corresponds to a local energy maximum due to its inherent instability, whereas distorted structures are energetically favored, as illustrated in Figure~\ref{fig3}. This behavior is well-documented in perovskite oxides, where cooperative J-T distortions lower the total energy and enhance the structural stability \cite{rondinelli2011structure}.

The comparison between the FM and PM states under the conditions of structural distortion further reveals intriguing stability trends. While both FM and PM states exhibit a preference for distorted structures over the ordered cubic phase, the FM state stabilizes the distorted structures more effectively. This observation is supported by prior theoretical investigations of transition metal perovskites, which indicate that magnetic ordering can influence the energy landscape by coupling to structural distortions \cite{mizokawa1995unrestricted,phelan2006nanomagnetic}. Furthermore, the stability of the magnetic state is found to be contingent upon the distortion angles, suggesting a sensitivity of the magnetic state to the structural parameters. This sensitivity of the magnetic state to structural distortions is further substantiated by Woicik $et\ al.$, who observed a change in the Curie temperature of LSC films under different strains \cite{woicik2011effect}.

Our analysis of oxygen vacancy formation energies also reveals a significant dependence on the structural distortion, cf.~Fig.~\ref{fig4}. When considering the most stable distorted structure, the vacancy formation energy is approximately 1.2~eV. While, if the ordered structure is used as a reference, the vacancy formation energy is drastically reduced to around 0.4~eV. Several previous studies on oxygen-deficient perovskites have emphasized the importance of lattice distortion too, showing that neglecting it leads to substantial underestimation of vacancy formation energies \cite{baldassarri2023accuracy, wang2024landscape}. More specifically, the research conducted by Baldassarri~$et~al.$~\cite{baldassarri2023accuracy} has indicated an energy underestimation of about 1 eV per oxygen vacancy for perovskite crystals, is in close agreement with our result. Consequently, our findings reinforce the need to incorporate structural distortions in defect calculations to obtain reliable energetic estimates.

We then analyzed the crystal structural distortion caused by the dopants. However, we did not find significant changes in bond angle and bond length during element substitution. Here we use the average bond angle because TM-O octahedra are always influenced by neighboring octahedra. We compared the effect of dopants on the bond angles for pristine crystals, crystals with V$_{\text{O}}$, and the same magnetic state, respectively. Compared to pure LSC, the average bond angle change in these cases for doped LSC is always in 1°. Therefore, we suggest that the energy changes due to crystal structure changes caused by these dopants can be neglected. Instead, these energy variations are attributable to the oxidizability of the dopants themselves, as well as to differences in response of the dopants to different magnetic states.

The difference of $E_{\text{V}}^{\text{O}}$ between the two magnetic states can be qualitatively explained by the double exchange interaction. The O atom between the two TM atoms is suggested to act as a bridge for double exchange in a FM perovskite oxide \cite{anderson1955considerations}. The presence of V$_{\text{O}}$ diminishes the bridge, thereby increasing the energy of the entire system. Consequently, the $E_{\text{V}}^{\text{O}}$ is expected to be higher in the FM state compared to the PM state. However, as illustrated in Figure~\ref{fig6}, the $E_{\text{V}}^{\text{O}}$ in FM and PM for Ni and Cu dopants are comparable, suggesting that Ni and Cu interrupt the double exchange before the presence of V$_{\text{O}}$. This indicates that the low spin of Ni and Cu dopants hinder the double exchange. A similar phenomenon was reported by Mizokawa~$et~al.$~\cite{mizokawa1995unrestricted}, who indicated that the low-spin Ni and Cu compounds in perovskites weaken the double exchange, which in turn destabilizes spin ordering.

The hypothesis of defect effects on the double exchange mechanism can also provide a qualitative framework for understanding their influence on the magnetic ordering temperature in LSC. Experimental studies by Baskar $et$~$al.$~\cite{baskar2008high} have shown that the presence of oxygen vacancies in LSC leads to a reduction in the magnetic ordering temperature compared to the pristine crystal. This observed decrease can be attributed to a reduction in the energy difference between the PM and FM states, a trend consistent with the results presented in Figure~\ref{fig7} (Co, Mn and Fe). Specifically, the suppression of the double-exchange mechanism, caused by oxygen vacancies, decreases the relative stability of the ferromagnetic state, thereby lowering the magnetic ordering temperature. In contrast, for low-spin dopants, such as Ni and Cu, the presence or absence of V$_{\text{O}}$ has little effect on the relative stability of the two magnetic states in the doped LSC. In this case, the roles of oxygen vacancies and low-spin dopants in hindering the double exchange interaction are consistent. %Consequently, our findings suggest that the sensitivity of the oxygen vacancy concentration to variations in the magnetic ordering temperature may be modulated by the choice of dopants in LSC, where the presence of V$_{\text{O}}$ does not affect the relative stability of the two magnetic states in the low spin TM doped LSC.

%Consequently, the change of band center distance between the two orbitals of the chemical bond is able to qualitatively analyze the 
To understand the different trends of double exchange interactions under various doping elements, we analyzed the electronic structure. The double exchange interaction exerts a significant influence on the distribution of electrons \cite{meetei2013theory,erten2013theory}, and this phenomenon can be observed in the pDOS. Stronger double exchange increases the degree of orbital overlap and the strength of chemical bond \cite{lv2010orbital}, which in turn affects the corresponding band center distances. In the energy range under consideration, the density of states is predominantly governed by contributions from the B-site TM 3$d$ orbitals (including both $t_{2g}$ and $e_g$ components) and O 2$p$ orbitals. The occupied state centers of the O 2$p$ and TM 3$d$ bands are labeled in Figure~\ref{fig8}. In this figure, we provide data for two magnetic states, FM and PM, for comparison. Note that the band center is calculated under the conditions of pristine crystals without V$_{\text{O}}$. We expect to correlate the variations between them with the trends of $E_{\text{V}}^{\text{O}}$ in different doping systems. %However, it is important to note that the band center distances are also influenced by the characters of the doping elements themselves. Therefore, we provide data about the two magnetic states, FM and PM, and quantify the double exchange effect of the different systems in terms of the change in the band center distances in the following.

\begin{figure}
	\centering
	\begin{overpic}[width=0.45\textwidth]{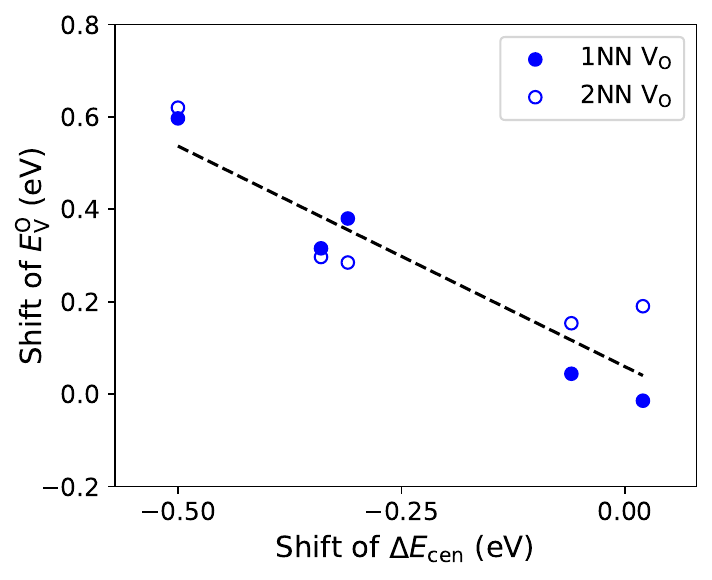}
	\end{overpic}
	\caption{Scatter plot of the energy difference between O 2$p$ and TM 3$d$ band centers versus the energy difference between FM and PM states in doped LSC. Solid and open blue circles denote the two different oxygen vacancies. The dashed line is the linear fitting of all data points.}
	\label{fig9}
\end{figure}

%Given that the doping element constitutes a partial obstruction to the double exchange, its effect is dependent on the degree of spin polarization. It is improbable that the doping element entirely obstructs the double exchange, while the presence of V$_{\text{O}}$ is likely to completely block the effect. Consequently, it can be deduced that a correlation exists between the variation in the band center distance and the change in the vacancy formation energy between the two magnetic states. Figure~\ref{fig9} illustrates this relationship. The observed relationship is approximately linear, thereby substantiating our prior hypothesis.

%Overall, the present study demonstrates that the oxygen vacancy formation energy of doped LSCs differs between the ground state FM and PM. This discrepancy cannot be quantitatively analyzed in terms of pristine crystal structure alteration and oxidizability of the doped elements. Utilizing the ground state FM to assess the oxygen vacancy formation energies of doped elements may lead to qualitative inaccuracies. For instance, the doping of Fe and Ni may result in a decrease in some oxygen vacancy formation energies in the FM state; however, these oxygen vacancy formation energies remain elevated in the PM state, cf.~Fig.~\ref{fig6}. Consequently, when assessing the impact of dopant elements, it is imperative to consider the influence of the magnetic state.
Figure~\ref{fig9} illustrates the relationship between the shift of $\Delta E_{\text{cen}}$ and the shift of $E_{\text{V}}^{\text{O}}$. These shifts arise from the difference between FM and PM states (FM minus PM). The band center takes into account the combined effects of the properties of the elements themselves and the double exchange interactions. We refrain to label the elements in this figure, as all systems should conform to our hypothesis. All the data points form an approximate linear relationship. 

This linear relationship confirms our earlier hypothesis that defects influence double exchange interactions. The doping elements partially obstruct the double exchange interaction, whereas the presence of V$_{\text{O}}$ is likely to completely obstruct them. When the doping elements fully obstruct the double exchange interactions, there is little change in the shift of $E_{\text{V}}^{\text{O}}$, corresponding to the right part of the line. Conversely, when the doping elements do not fully hinder double exchange interactions, the V$_{\text{O}}$ takes responsibility for hindering the remaining double exchange interactions, thus increasing $E_{\text{V}}^{\text{O}}$, which corresponds to the left part of the line. Therefore, the influence of doping elements on $E_{\text{V}}^{\text{O}}$ exhibits inconsistent trends under different magnetic states, and this inconsistency is determined by the varying degrees of hindrance that defects impose on double exchange interactions.

\section{Summary\protect} \label{summary}

In this study, we utilized DFT calculations to systematically explore the effects of 3$d$ transition-metal doping on the oxygen vacancy formation energies in the LSC perovskite material.  Two magnetic orders were considered to uncover the intricate relationships among dopant chemistry, magnetic order, and defect energetics. Our results reveal that the formation energies of oxygen vacancies in the doped LSC are strongly influenced by both the magnetic state of the system and the choice of 3$d$ transition-metal dopant. The PM state, which serves as a model for general operational conditions, exhibits a different trend of oxygen vacancy formation energies when compared to the FM ground state. This difference cannot be quantitatively analyzed in terms of pristine crystal structure alteration and oxidizability of the doped elements. Utilizing the ground state FM to assess the oxygen vacancy formation energies of a doped system may lead to qualitative inaccuracies. The reason for this inaccuracy is the effect of the doping element on the double exchange interaction, as we discussed in Fig.~\ref{fig9}. This emphasizes the need to account for magnetic disorder in computational studies. These findings lay the foundation for the development of customized perovskite materials with enhanced performance in electrochemical applications, such as SOFCs and oxygen separation membranes, where precise control over oxygen vacancy concentrations is paramount.

\begin{acknowledgments}
This work was funded by the European Union under Grant Agreement 101080142, as part of the EQUALITY project. The authors thank David David (Capgemini SE) for fruitful discussions. The work was enabled in part by the Nextsim HPC at Chemnitz University of Technology. 
\end{acknowledgments}

%\nocite{*}
\nocite{}

\bibliography{lsc_bulk_v5}
% Produces the bibliography via BibTeX.

\end{document}